\documentclass[prl,showpacs,,amssymb,twocolumn]{revtex4}

\newcommand{\be}{\begin{equation}}
\newcommand{\ee}{\end{equation}}
\newcommand{\la}{\label}

\newcommand{\C}{\mathbb{C}}
\newcommand{\R}{\mathbb{R}}

\def\p{\partial}

\usepackage{graphicx}
\usepackage{bm}
\usepackage{amssymb}

\begin{document}

\title{Optimal approximation of harmonic growth clusters by orthogonal polynomials}

\author{Ferenc  \surname{Balogh}$^{{\sharp},{\ddagger}}$}
\author{Razvan  \surname{Teodorescu}$^{\sharp}$}
\affiliation{$^{\sharp}$Center for Nonlinear Studies, LANL}
\affiliation{$^{\ddagger}$Concordia University, Montr\'eal}
\date{\today}

\begin{abstract}
Interface dynamics in two-dimensional systems with a maximal number of conservation
laws gives an accurate theoretical model for many physical processes, from the 
hydrodynamics of immiscible, viscous flows (zero surface-tension limit of Hele-Shaw
flows, \cite{Hele-Shaw}), to the granular dynamics of hard spheres \cite{Jag-Nag}, 
and even diffusion-limited aggregation \cite{Halsey}. Although a complete solution 
for the continuum case exists \cite{Krichever-Mineev-Weinstein_et_al, MWZ}, efficient 
approximations of the boundary evolution are very useful due to their practical applications 
\cite{Put02}. In this article,  the approximation scheme based on orthogonal polynomials 
with a deformed Gaussian kernel \cite{Teodorescu04} is discussed, as well as relations  
to potential theory.
\end{abstract}

\pacs{05.70.-a, 02.10.Yn, 02.30.Tb, 05.30-d, 05.40.-a, 05.50.+q}

\maketitle

\paragraph*{Introduction --} A large class of two-dimensional processes, both deterministic
and stochastic, have been mapped to a powerful mathematical model known as {\emph{harmonic}} 
(or {\emph{Laplacian}}) {\emph{growth}} \cite{Bear, CahnHill, DLA81, Gollub, Glicksman, Howison86, KruSeg, Langer, Nakaya}. In this model, a domain $D_+(t_0)$ {\emph{grows}} in time 
$t$, i.e. its normalized area $t_0 = \pi^{-1}\int_{D_+ }dxdy$ (areas in units of $\pi$ throughout) has a linear dependence  $t_0=Qt$, with $Q$ the {\emph{pumping rate}}. 

This prescription is obtained as a consequence of the complete formulation as an {\emph{exterior}} problem: if $\Gamma$ represents the boundary of $D_+$ and $D_-$ its complement, we seek the harmonic field $p$ (which may represent pressure, concentration, etc) and the velocity field $\vec{v}$, such that:
\begin{equation} \la{eq} 
\left \{
\begin{array}{lcl}
v_n = -\p_n p, & p = 0 & \mbox{on} \,\, \Gamma, \\
\vec{v} = - \vec \nabla p, & \Delta p = 0 & \mbox{on} \,\, D_-,  \\
p \sim - \log |z|, & \,\, z \to \infty & \in D_-
\end{array}
\right.
\end{equation}
with $v_n, \p_n p$ their normal components on $\Gamma$, respectively.

\begin{figure}[htb]
\begin{center}
\includegraphics*[height=3cm]{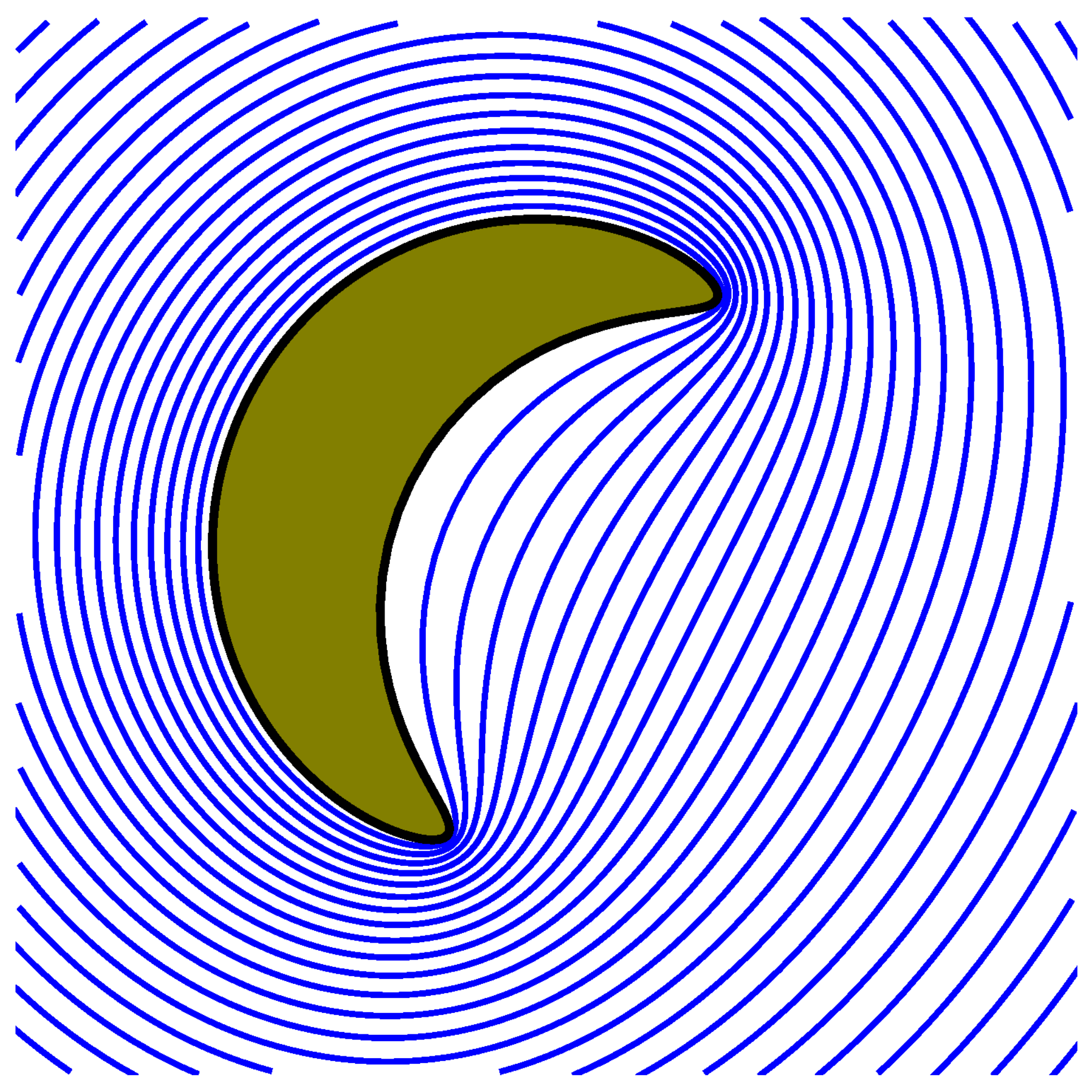} \ \ \includegraphics*[width=3cm]{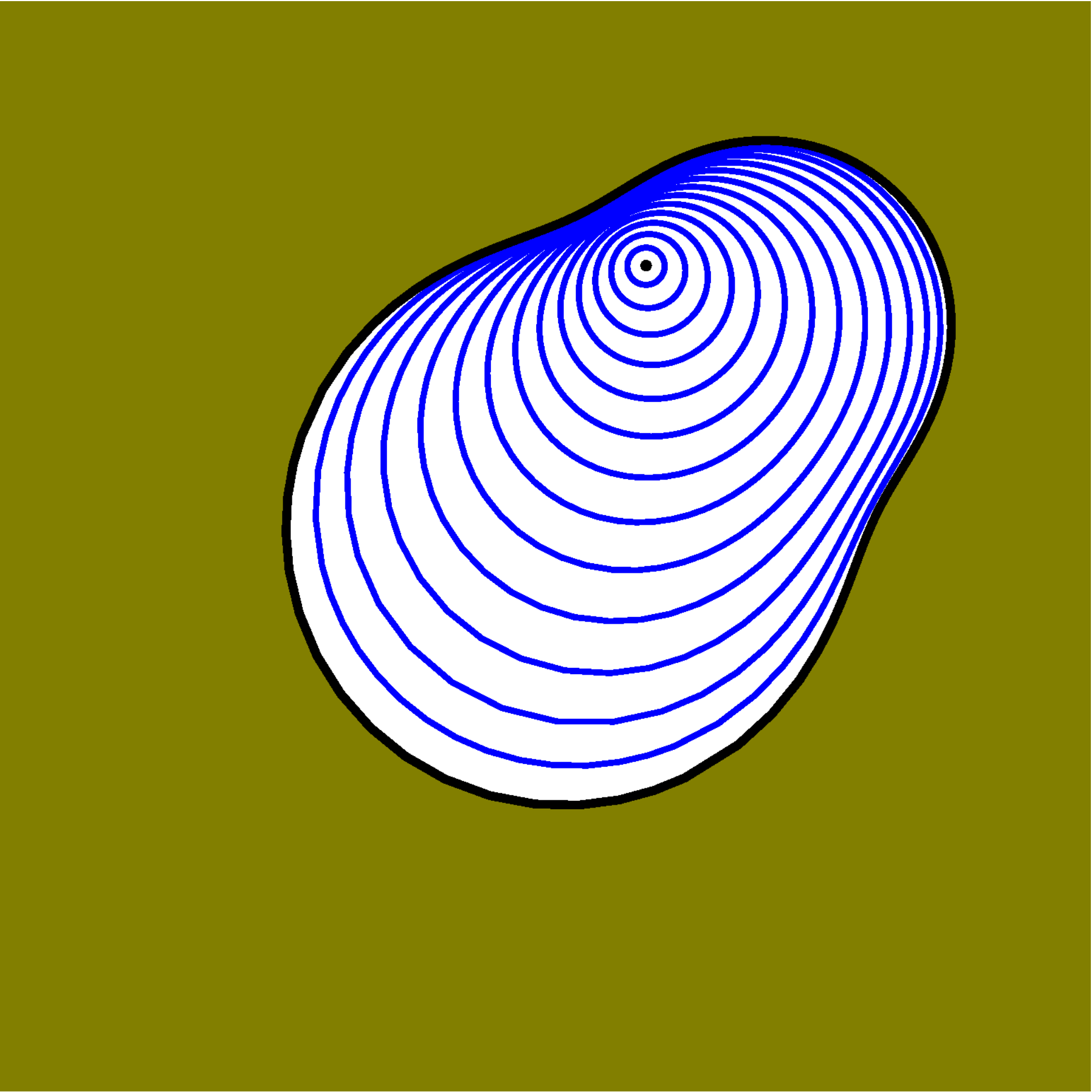}
\caption{Simply connected droplets and equipotential lines for the interior and the exterior domains.}
\label{droplets}
\end{center}
\end{figure}

An equivalent formulation of the {\emph{interior}} problem was studied by S. Richardson \cite{Richardson72}, where the roles of exterior and interior domains are interchanged: $p$ is harmonic inside up to a finite number of isolated logarithmic singularities (\emph{sources} and \emph{sinks}).
Assume for simplicity a single point source at the origin $z=0$ with pumping rate $Q$, or, in terms of the pressure $p$, $ \Delta p = \frac{Q}{2\pi}\,\delta^2(z)$. Richardson proved that for any function $h \in L^1(D_+)$ harmonic in a neighborhood of $D_+$ we have
\be
\frac{d \langle h \rangle_+ }{d t} = Q h(0),
\ee
where $\langle h \rangle_{+}= \int_{D_+(t)} h(z) dxdy$. Therefore, taking $h(z)=z^k$ we have that the \emph{interior harmonic moments}
\begin{equation}
v_k = \frac{1}{\pi}\int_{D_+(t)}\,z^{k}\,dx\,dy, \qquad k=1,2,\ldots
\end{equation}
are preserved while the area grows linearly with $t$.

Mapping the interior problem by means of a simple coordinate change $\tilde z = 1/z$ and changing the source to a sink ($Q \to -Q$) implies that in the original exterior problem for $D_-$ , $0 \in D_-$, the \emph{exterior harmonic moments}
\begin{equation}
t_k = -\frac{1}{k \pi}\int_{D_-(t)}\,z^{-k}\,dx\,dy, \qquad k=1,2,\ldots
\end{equation}
(the interior moments of $D_-$ now) are preserved by the evolution and $dt_0 / dt = Q$ \cite{Varchenko/Etingof:1992}, and we set $Q=1$. 

For a simply connected droplet $D_+$, the solution of the problem is easily expressed through the conformal map taking the exterior of the unit disk $|\zeta| >  1$ univalently onto $D_-$, and matching the points at infinity,
\be \la{map}
z = f(\zeta) \equiv r\zeta + \sum_{k = 0}^{\infty} u_k \zeta^{-k}, \quad r >0,
\ee
with respect to which the pressure and normal component of boundary velocity read
\be
p(z) = - \log|f^{-1}(z)|, \quad v_n = |\p_z f^{-1}(z)|.
\ee

As shown in \cite{MWZ}, conservation of exterior harmonic moments is equivalent to the statement that there is a canonical transformation from the variables $z, z^{\sharp}$,
\be
z = f(\zeta), \quad 
z^{\sharp} = \overline{f}(\zeta^{-1}),
\ee
to the variables $t_0, \log \zeta$, i.e. $
\zeta \left [\frac{\p z}{\p \zeta}\frac{\p z^{\sharp}}{ \p t_0} -  
\frac{z^{\sharp}}{\p \zeta}\frac{\p z}{\p t_0} \right ] = 1.
$
We note the alternative formulation 
$ d z \wedge d z^{\sharp} = d \log \zeta \wedge d t_0.$

In this paper, we describe an optimal approximation procedure for the boundary 
$\Gamma(t)$, as well as for the Cauchy transform of $D_+$, 
\be
C_{D_+} (z) = \frac{1}{\pi}\int_{D_+}\frac{dx' dy'}{z-z'}, \quad z'=x'+iy', \,\,
 z \in D_-.
\label{cauchy}
\ee
\paragraph*{An example --}
A superficial analysis of the model described here would lead to the conclusion that, given 
the integrability of the system, specifying a set of exterior harmonic moments $\{ t_k \}$ at 
a given, initial area $T_0$, should yield a solution for $t_0 > T_0$ without major analytical 
difficulties. In reality, reconstructing the conformal map (\ref{map}) which corresponds to the
data $T_0, \{ t_k \}$ is typically a very difficult problem. There are few known correpondences,
between classes of moments $\{ t_k \} $ and classes of shapes (\ref{map}) (see \cite{Teodorescu04} for examples). An illustrative case occurs when the exterior harmonic moments form a simple sequence
\be
\label{geometric}
t_k  = -\frac{\beta}{k}a^{-k} \qquad k = 1, 2, \dots,
\ee
where $\beta \in \R^{+}$ and $a\in \C\setminus \{0\}$ are given parameters. Defining two radii
\be
R_1 = \sqrt{\beta}, \qquad R_2 =\sqrt{\frac{1+2\beta}{2}},
\ee
we have two completely different cases: if $|a|+R_1 \leq R_2$ then we have a doubly connected domain bounded by circles and for 
$|a|+R_1 > R_2$ the domain is simply connected given by an exterior conformal map of the form
\be \la{joukowski}
f(\zeta)  = r\zeta + u +\frac{v}{\zeta - A}, \qquad r>0,\,\, |A|<1,
\ee
(see FIG.~\ref{maps}). The parameters of $f(\zeta)$
are related to the deformation parameters as follows: setting $u=v/A$,
\be\la{corr}
\left \{
\begin{array}{rcl}
      \displaystyle \beta  & = & \displaystyle t_0 - r^2 +\frac{r\bar v}{{\bar A}^2},  \\
     a & = &  \displaystyle \frac{r}{\bar A} + \frac{v}{A}  + \frac{v \bar A}{1-|A|^2},
\end{array}\right.
\ee
where $t_0 = r^2 - |v|^2/(1-|A|^2)^2$ is the area. If $|a|+R_1 > R_2$ then
the above equations have a unique solution for $r, v$ and $A$ in terms of $\beta$ and $a$. 

\begin{figure}[htb] 
\begin{center}
\includegraphics*[width=0.48\textwidth]{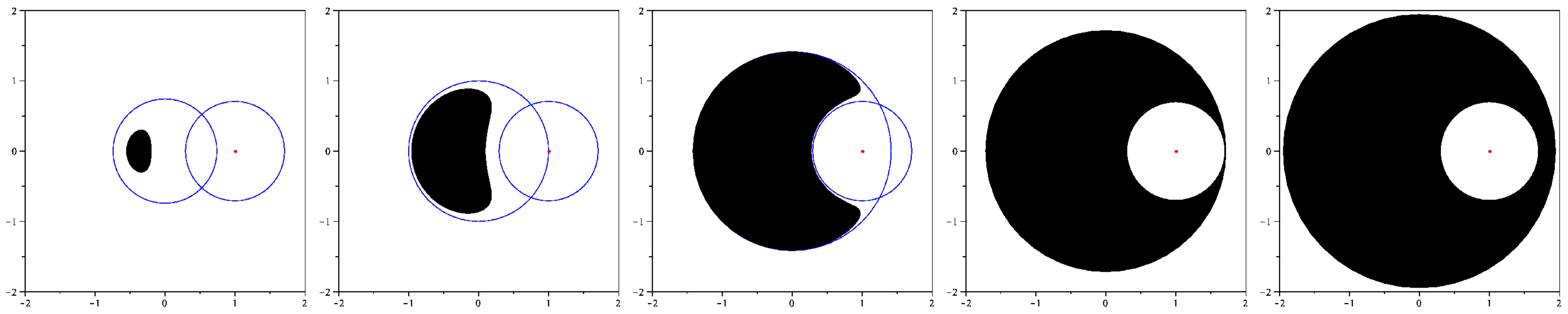}
\caption{Laplacian growth for the exterior moments (\ref{geometric})
.}
\label{maps}
\end{center}
\end{figure}


\paragraph*{Domain approximation via orthogonal polynomials--}
The fact that, even for a rather simple shape (\ref{joukowski})
the correspondence  (\ref{corr}) is quite intricate shows the need for, 
and practical value of, efficient approximation methods. To that 
end, we define the following family of orthogonal polynomials: 
consider the domain specified by the exterior harmonic moments 
$\{ t_k \}$ and $t_0$. The function defined in a neighborhood of the origin by
\be
V(z) = \sum_{k \ge 1} t_k z^k, \quad z \to 0, 
\ee
is preserved by the harmonic growth. Now consider the function $W(z) = |z|^2 - 2 \Re V(z)$, which we label {\emph{confining potential}}, and suppose that
$$
\int_{\C} |z|^n e^{-N W(z)} d^2 z < \infty \qquad n =0,1,2,\dots
$$
for all for values of the scaling parameter $N>0$.
For fixed $N$, the orthogonal polynomials $\{ P^{(N)}_n(z) \}$ of the weight function $e^{-N W(z)}$ are 
defined by
\be
\int_{\mathbb{C}} P^{(N)}_n(z) \overline{P^{(N)}_m(z)} e^{-N W(z)} d^2 z = 
\delta_{nm}. 
\ee
The approximation method presented in this work is based on the following 
statement: as 
\be
\label{scaling_limit}
n \to \infty, \quad N \to \infty, \quad \frac{n}{N} \to t_0,
\ee
the weighted polynomials $|P^{(N)}_n(z)|^2 e^{-NW(z)}$ (which we denote by $\rho_n^{(N)}(z)$ 
in the following) converge to the conformal measure of the domain $D_+(t)$, 
with support  $\Gamma(t)$:
\be \la{limit}
\rho_n^{(N)}(z) = 
|\p_z f^{-1}(z)|  \left  [ 1+ O \left ( \frac{1}{N} \right ) \right ],
\,\, n, N \to \infty. 
\ee
The proof of this result appeared first in \cite{Teodorescu04}. We do 
not repeat the entire argument, as it would require too much space, 
but recollect the main ideas: starting from the differential equations
satisfied by the weighted functions $\psi_n(z) = P_n^{(N)}(z) e^{-N W(z)/2}$, 
with respect to variables $n/N$ and $z$, we integrate perturbatively 
in powers of $N^{-1}$, and obtain the expression (\cite{Teodorescu04}, 
equation (76)):
$$
\psi(z) \sim \sqrt{[f^{-1}(z)]'} \exp \left [N \left (-\frac{|z|^2}{2} + \Re \int^z S(\zeta) d\zeta \right ) \right ],
$$
where the {\emph{Schwarz function}} $S(z)$ is defined by the identity 
$S(z) = \bar z, z \in \Gamma$. It is known to have the expansion 
\be
S(z) = V'(z) + \frac{1}{\pi}\int_{D_+}\frac{dx' dy'}{z - z'}, \quad z \to \infty
\ee
Since the exponent in the asymptotic expression of $\psi(z)$
 vanishes on the boundary $\Gamma$ and gives a Gaussian decay away 
 from it, the weighted polynomials  $\rho_n^{(N)}(z)$ are 
 described, in the $n, N \to \infty$ limit, by the conformal measure (\ref{limit}). 
 However, this asymptotic result says very little about the behavior of 
 $\rho_n^{(N)}(z)$ for {\emph{finite}} values of $n, N$. 

In the remainder of this paper, we present numerical evidence for the convergence 
properties of $\rho_n^{(N)}(z)$ at finite values of their order. We show that the 
agreement between $\rho_n^{(N)}(z)$ and $|[f^{-1}(z)]'|$ is excellent for values
of $n$ as little as $n=20$, and present potential applications of this property. 
One obvious consequence is related to reconstruction algorithms of domains 
in this class: assume that the Schwarz function of the domain $D_+$ has a branch
cut on $\gamma \in D_+$, with real, positive jump function $\rho_s(z)$.  Obviously, 
this may include the case of meromorphic functions with  poles in $z_p \in D_+$, for 
which $\rho_s(z) = \sum_p a_p \delta(z-z_p)$. Then from the asymptotic result
\be
\frac{1}{N}\log |\rho_N^{(N)}(z)| \to W(z) + \frac{1}{\pi} \int_{D_+}
\log |z-z'|^2 dx' dy',
\ee
it follows immediately that the asymptotic distribution of zeros of the orthogonal
polynomials converges to $\rho_s(z)$:
\be \la{cut}
\lim_{N \to \infty} \frac{1}{N}\sum_{i=1}^N \log |z-z_i|^2
= \int_{\gamma}\rho_s(z') \log|z-z'|^2 dz'.
\ee

In a forthcoming publication \cite{Ed_Ferenc_Razvan}, we will give 
detailed proofs of (\ref{limit}, \ref{cut}). In this Letter, we provide 
a thorough numerical analysis supporting the asymptotic results.

\paragraph*{Simulations and numerical study --}

\begin{figure}[htb] \begin{center}
\includegraphics*[width=5cm]{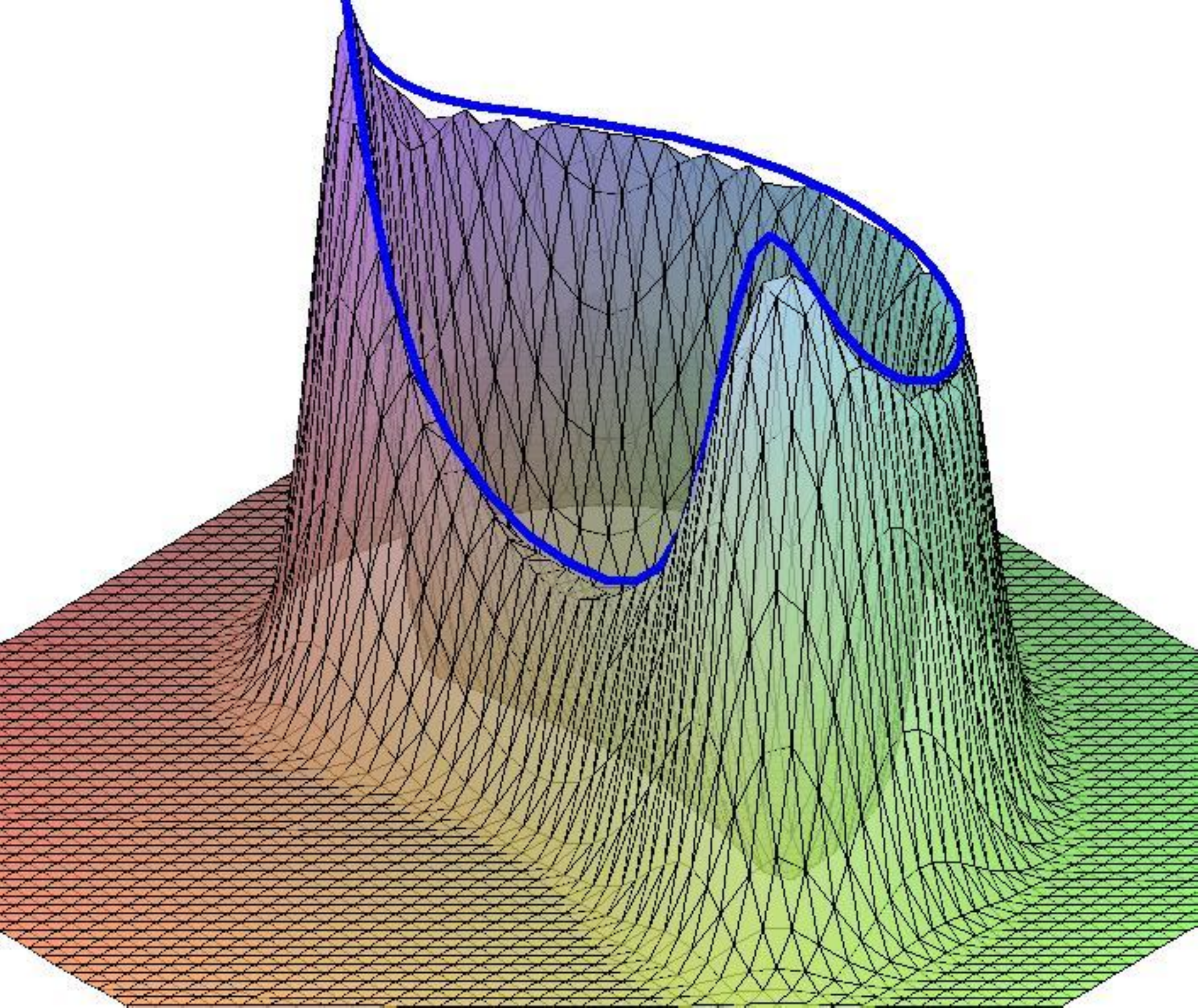}
\caption{Localization for the density for $n=20$ and the conformal measure.}
\label{density_vs_conformal}
\end{center}
\end{figure}

Let $t_0$ and the exterior harmonic moments be given through the potential $V(z)$.
To fix the scaling limit, let $N(n)= n/t_0$. For fixed $n$, we have to calculate the entries of the Gram matrix 
\be
g^{(n)}_{ij} = \int_{\C}z^i {\bar z}^j e^{-n/t_0 W(z)}d^2 z \quad i,j =0,\dots, n.
\ee  
For potentials $W(z)$ that are converging rapidly enough to infinity as $|z| \to \infty$, the exponentially decaying weight makes the planar numerical integration a feasible task. The stabilized Gram-Schmidt Algorithm provides the orthogonal polynomials $P_{n,N}(z)$, which is known to be very sensitive to the accuracy of the Gram matrix and thus requires very precise computation of  $\{ g^{(n)}_{ij} \}$ . Then  the density $\rho_n^{(N)}(z)$ is obtained from the polynomial $P^{(N)}_n(z)$.

Of course, the usefulness of this approximation scheme relies on the rapidity of the convergence in (\ref{limit}), which may not seem to be very promising.
However, our numerical experiment (FIG.~\ref{density_vs_conformal}) shows that in the example (\ref{geometric}) above the ``shape" of the conformal measure (the blue curve) is recovered very accurately by the weighted polynomial density of a degree as low as $n=20$.

\begin{figure}[htb] \begin{center}
\includegraphics*[width=4.3cm]{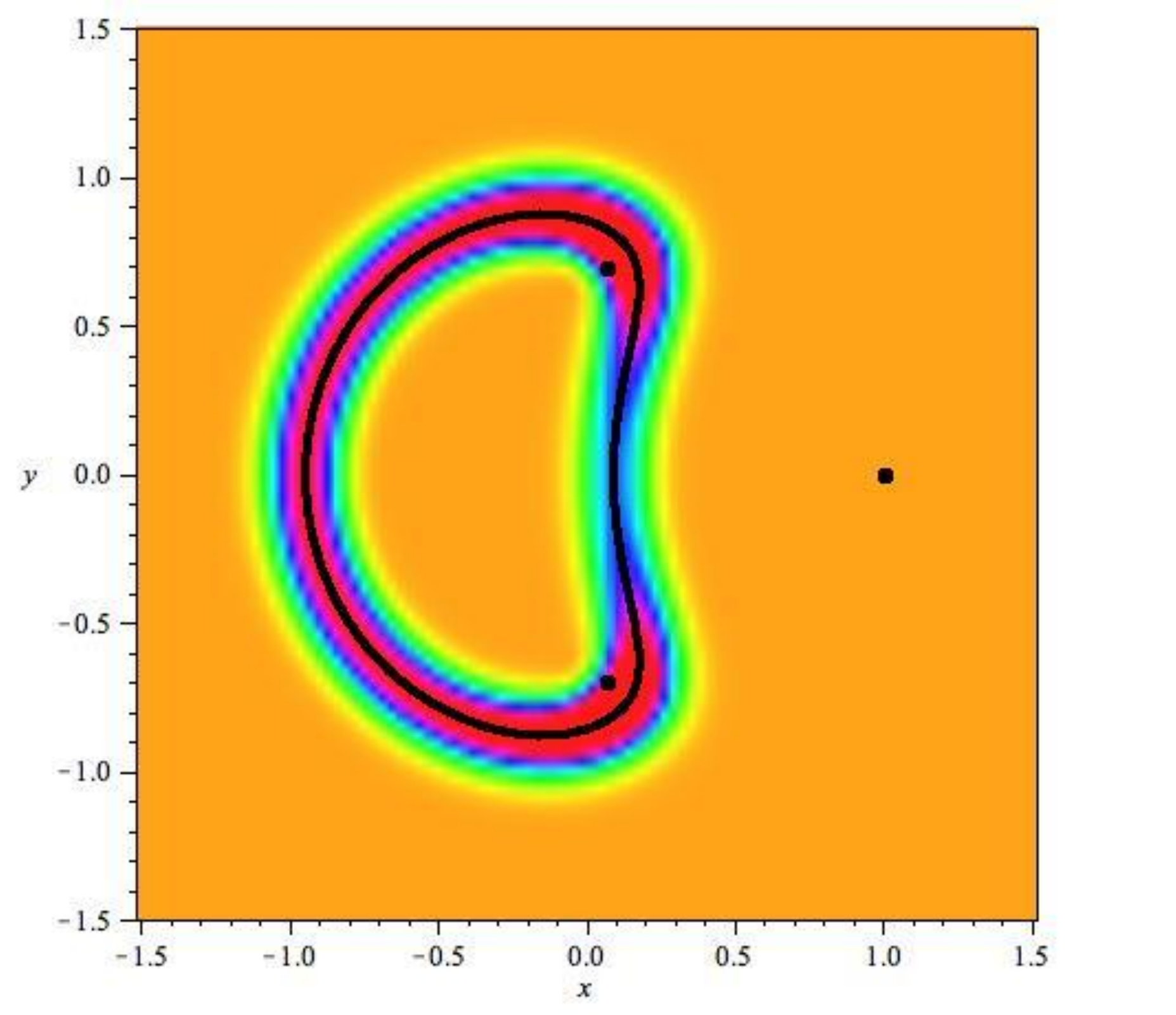} \ \includegraphics*[width=4cm]{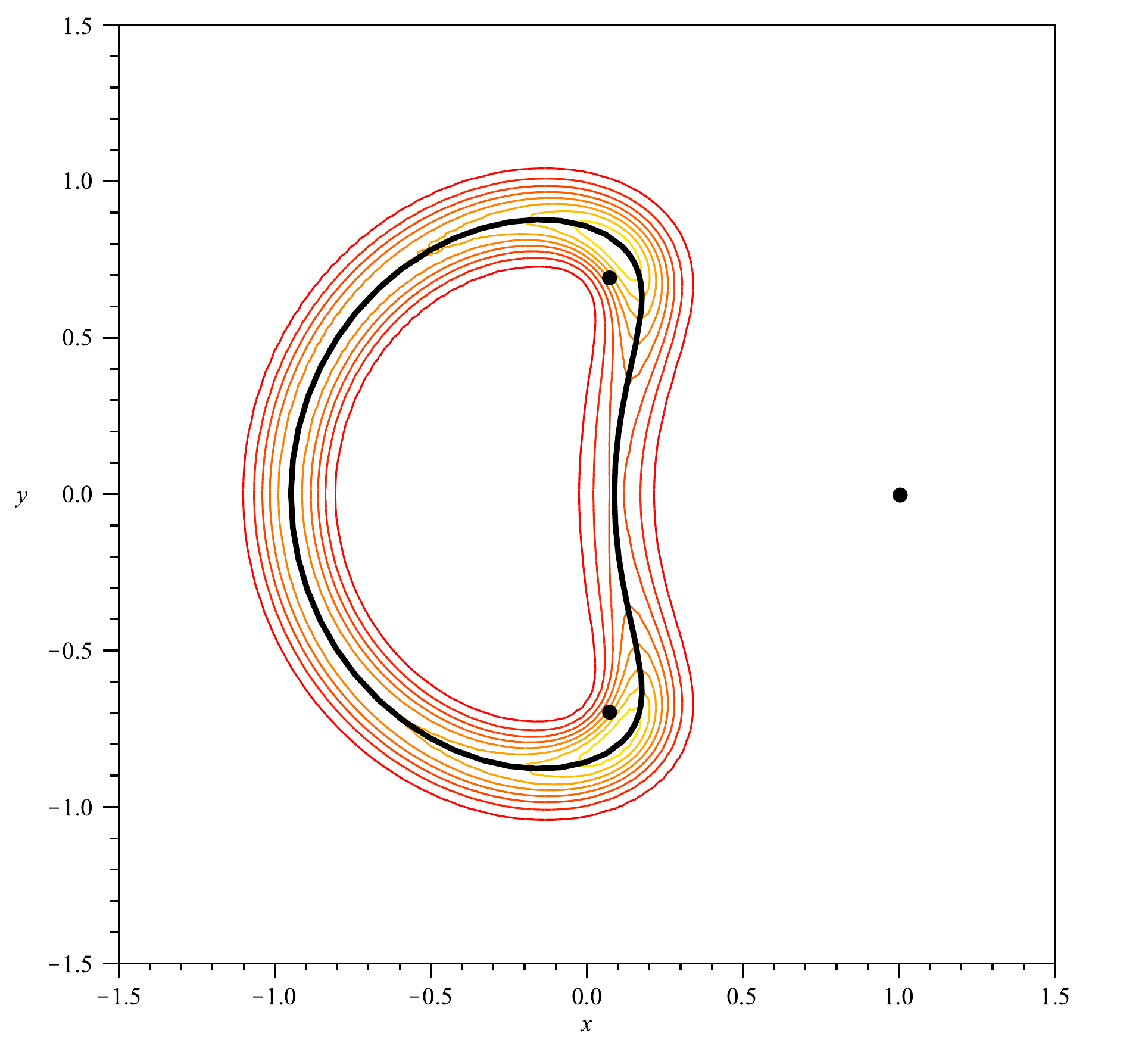}
\caption{Density plot and contour plot of the localized density}
\label{droplets}
\end{center}
\end{figure}

To illustrate the speed of convergence in (\ref{limit}), the Kullback-Leibler divergence or relative entropy of ${\rho_n^{(N)}(z)}$ with respect to the harmonic measure, given by
\be
\frac{1}{2\pi}\int_{0}^{2\pi} \left(\log{\frac{1}{|f'(e^{i\theta})|}} - \log{\rho_n^{(N)}(f(e^{i\theta}))}\right)d\theta,
\ee
 is calculated and plotted on the figure below.
\begin{figure}[htb] 
\begin{center}
\includegraphics*[width=0.48\textwidth]{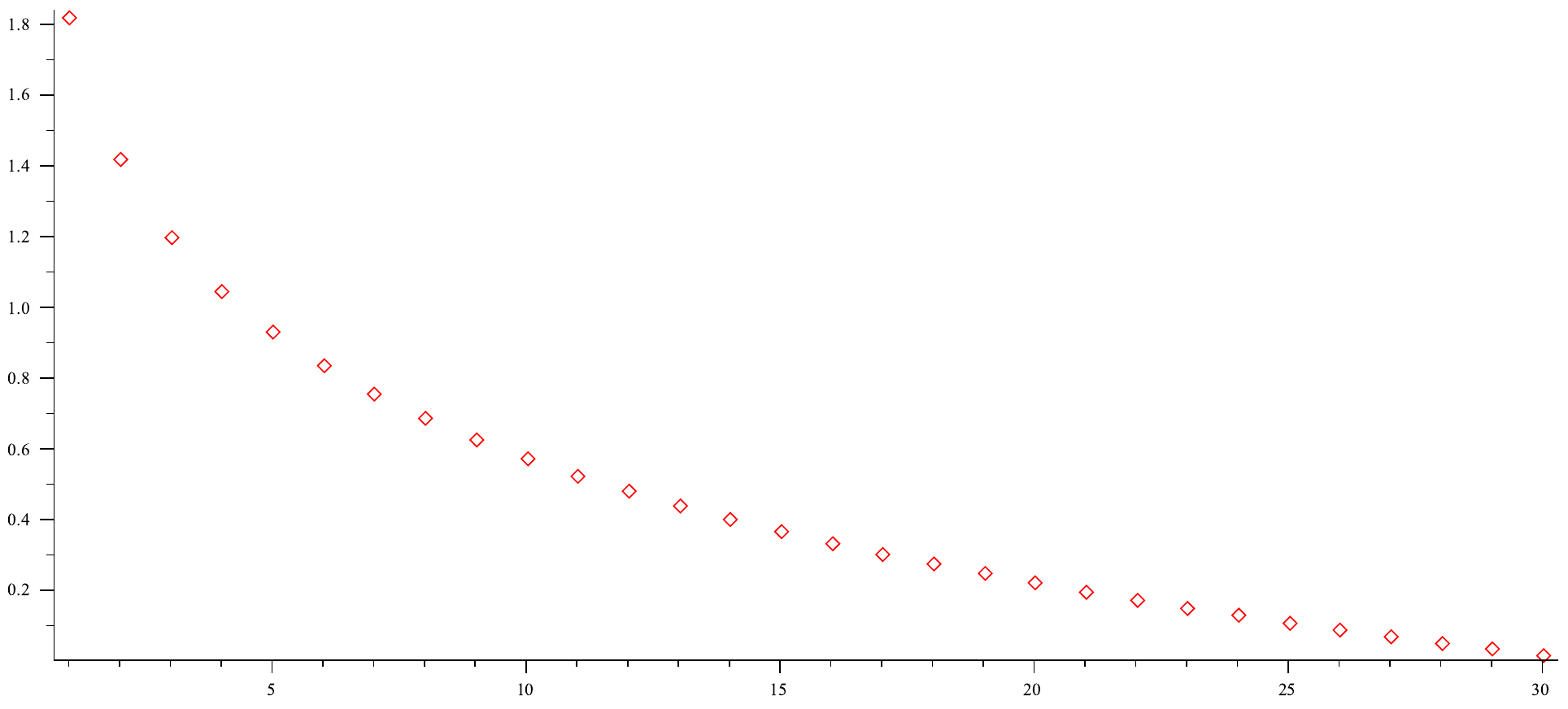}
\caption{The Kullback-Leibler divergence of ${\rho_n^{(N)}(z)}$ with respect to the harmonic measure.}
\label{maps}
\end{center}
\end{figure}

\begin{figure}[htb] \begin{center}
\includegraphics*[width=5cm]{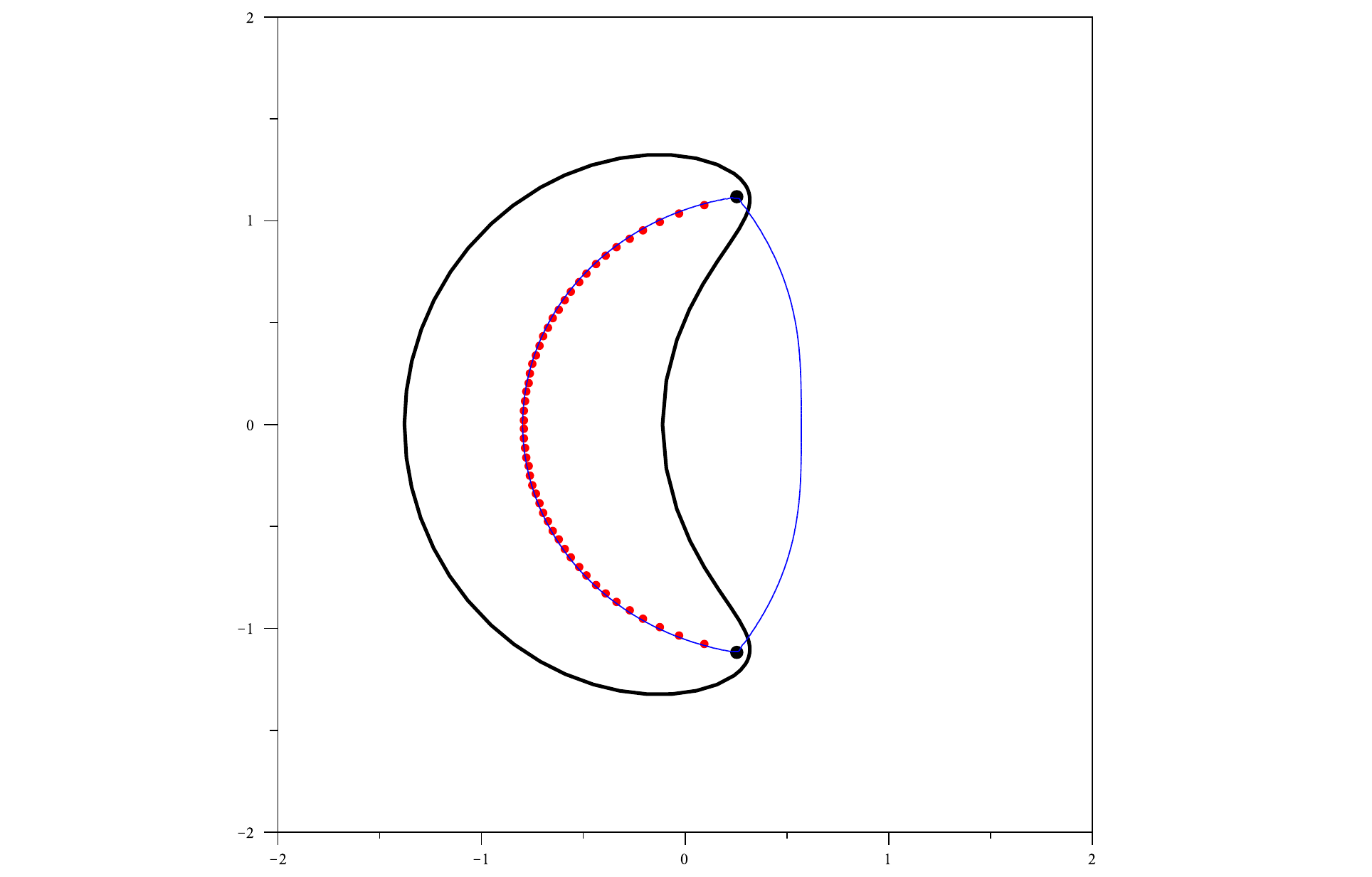} 
\caption{The critical trajectory and the zeroes for $n=50$.}
\label{trajectory}
\end{center}
\end{figure}

The asymptotic behaviour of the zeroes of orthogonal polynomials in the scaling limit (\ref{scaling_limit}) was also investigated in the particular case (\ref{geometric}). Since $f(\zeta)$ is a rational function of order two, the Cauchy transform $C_{D_+} (z)$ (\ref{cauchy})  in the exterior domain $D_-$ satisfies a quadratic equation 
$$
A(z)y^2 +B(z)y + C(z) =0,
$$
with rational coefficients in $z$ depending on the parameters of $f(\zeta)$.  As an algebraic function, $C_{D_+} (z)$ can be analytically continued on a plane with a branch cut connecting up the branchpoints
\be
z_{1,2} = \frac{v}{A}+Ar \pm 2\sqrt{rv}  
\ee
of the inverse mapping $f^{-1}(z)$. This ``conjugate electric field" created by the uniformly charged domain $D_+$ is mimicked by the field generated by the  normalized counting measure of the zeroes. However, these points seem to accumulate along some curve (as opposed to the equilibrium configuration in the presence of the background potential $W(z)$ -- the so-called \emph{Fekete points} -- which are distributed asymptotically uniformly). Since the asymptotic zero distribution must be real and positive, the natural choice is dictated by the \emph{Sokhotski-Plemelj} formula: the critical trajectory $\gamma$ is selected by the condition that the jump between the two solutions $y_{\pm} = (-B \pm \sqrt{B^2-4AC})/2A$ satisfies
\be
\Re\left((y_+(z)-y_-(z))dz\right) =0.
\ee  
The critical trajectory can be found by calculating
\be
\Phi(z) := \Re\left [\int_{z_1}^{z} (y_+(w)-y_-(w))dw\right ]
\ee
and then plotting the contour $\Phi(z) = 0$. Three trajectories are emanating from each branchpoint: there are two trajectories that connect $z_1$ and $z_2$, and 
the one contained by the domain attracts the roots. As can be seen in FIG.~\ref{trajectory}, the 
distribution of zeros (for $n=50$) and the trajectory $\gamma$ are almost indistinguishable. 

\paragraph*{Applications --} The method presented in this 
Letter allows to construct optimal approximations with high
convergence rates for either the boundary or the branch 
cuts characterizing domains from the harmonic growth class.
This may be used in a number of different situations; here 
we discuss two relevant examples: 

$(i)$ an outstanding problem in viscous two-dimensional 
flows is formation of boundary singularities (cusps). They 
are known to occur for finite values of the normalized area
$t_0$, and for many initial conditions \cite{Hohlov-Howison94}.
For a particular class of such cusps, with local geometry 
given by the scaling $x^2 \sim y^{2k+1}, k = 1, 2, \ldots$, 
it is not possible to continue the evolution of the boundary 
 beyond the cusp formation, and a weaker type of solution 
 is required. 
 
\begin{figure} \begin{center}
\includegraphics*[width=4cm]{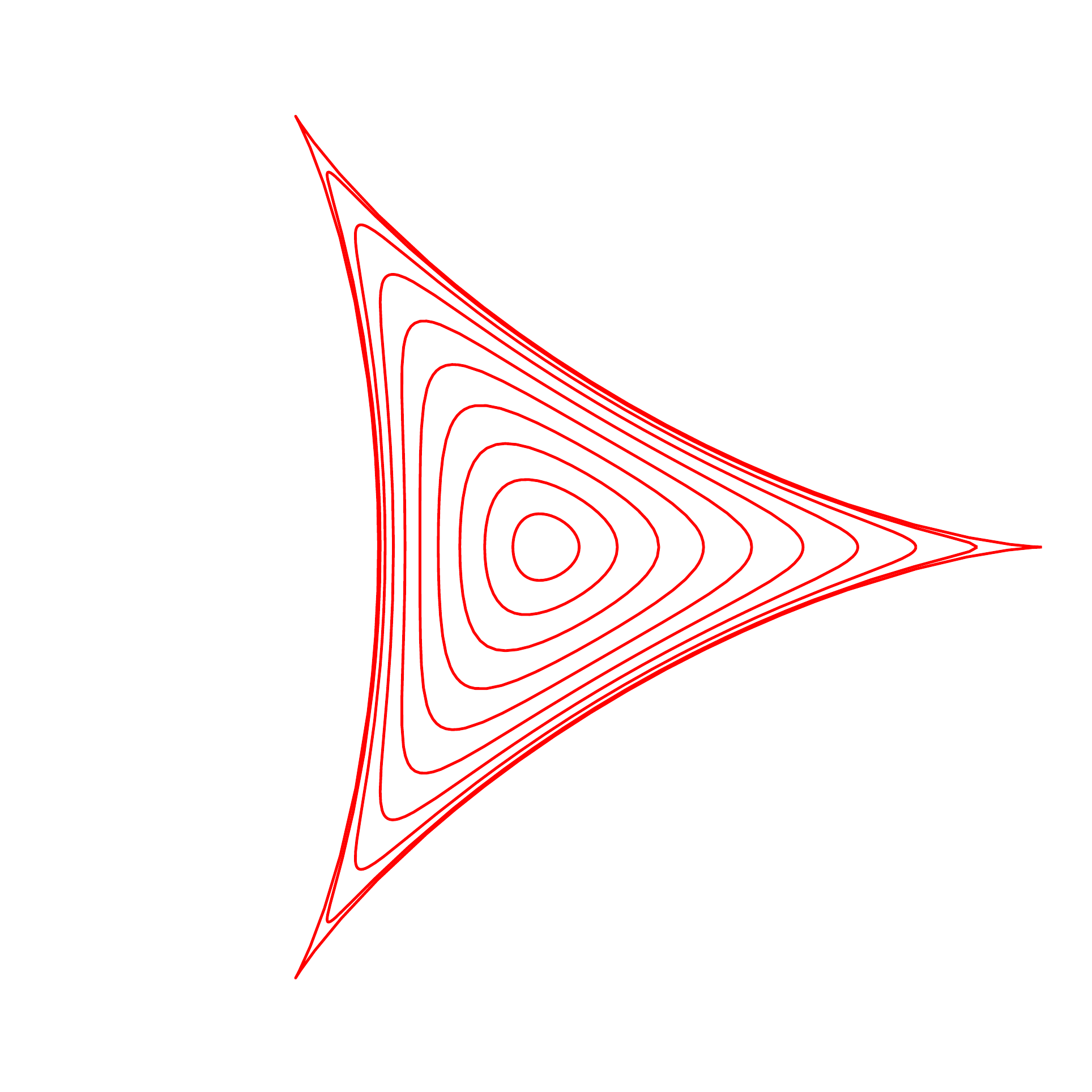}
\caption{Singular shape as a result of Laplacian growth}
\label{singular}
\end{center}
\end{figure}

The weak solution \cite{WT} is based on the 
equivalence between the distribution of zeros of the 
orthogonal polynomials and the branch cut of the Schwarz 
function, (\ref{cut}).  These two distributions generate the same
Newtonian potential in $D_-$ as the uniform distribution on
$D_+$ (real droplet), so they may be considered as equivalent solutions
before singularity formation, FIG.~\ref{singular}. However, after a cusp is formed, 
smooth (uniform) solutions are not possible anymore, while 
the distribution of zeros of the polynomials remains well-defined.
The conjecture is that this weak formulation will produce 
solutions which explain the famous fingering patterns observed
in physical realizations of this model. This 
will be shown in a forthcoming publication, 
where the algorithm will be used to 
construct numerical solutions according to this prescription, 
and compare with real, physical patterns \cite{Sw}.

$(ii)$ another practical application of the results 
presented in this Letter is an efficient algorithm for shape 
(boundary) reconstruction when the domain $D_+$ is given
through the reduced data $t_0, \{ t_k \}$. For example, such 
(reduced) representations arise in satellite imaging data 
compression \cite{Put02}. Shape reconstruction algorithms are 
then needed to find the boundary $\Gamma$, given the set of
moments $t_0, \{ t_k\}$, with particular emphasis on good convergence
rates. Since the data $t_0, \{ t_k\}$ is sufficient for constructing 
the family of orthogonal polynomials $P^{(N)}_n$ introduced here, we 
have a boundary approximation algorithm which gives excellent
results already at $n=20$.

\paragraph*{Acknowledgments --} 

This work was carried out under the auspices of the National Nuclear Security 
Administration of the U.S. Department of Energy at Los Alamos National 
Laboratory under Contract No. DE-AC52-06NA25396. R.T. acknowledges 
support from the  LDRD Directed Research grant on {\it Physics of Algorithms}.


\def\cprime{$'$} \def\cprime{$'$} \def\cprime{$'$} \def\cprime{$'$}
  \def\cprime{$'$}


\begin{thebibliography}{10}

\bibitem{Hele-Shaw}
H.~S.~S. Hele-Shaw.
\newblock {\em Nature}, 58(1489):34--36, 1898.

\bibitem{Jag-Nag}
X.~{Cheng}, L.~{Xu}, A.~{Patterson}, H.~M. {Jaeger}, and S.~R. {Nagel}.
\newblock {\em Nature Physics}, 4:234--237, March 2008.

\bibitem{Halsey}
T.~C. {Halsey}.
\newblock {\em Physics Today}, 53:36--41, 2000.

\bibitem{Krichever-Mineev-Weinstein_et_al}
I.~Krichever, M.~Mineev-Weinstein, P.~Wiegmann, and A.~Zabrodin.
\newblock {\em Physica D}, 198(1-2):1--28, 2004.

\bibitem{MWZ}
M.~Mineev-Weinstein, P.B. Wiegmann, and A.~Zabrodin.
\newblock {\em Physical Review Letters}, 84:5106, 2000.

\bibitem{Put02}
M.~Putinar.
\newblock {\em Numer. Math.}, 93(1):131--152, 2002.

\bibitem{Teodorescu04}
R.~Teodorescu, E.~Bettelheim, O.~Agam, A.~Zabrodin, and P.~Wiegmann.
\newblock {\em Nuclear Phys. B}, 704(3):407--444, 2005.

\bibitem{Bear}
J.~Bear.
\newblock {\em Dynamics of fluids in porous media}.
\newblock Elsevier (New York), 1972.

\bibitem{CahnHill}
J.~W. Cahn and J.~E. Hilliard.
\newblock {\em The Journal of Chemical Physics}, 28(2):258--267, 1958.

\bibitem{DLA81}
T.~A. Witten and L.~M. Sander.
\newblock {\em Phys. Rev. Lett.}, 47(19):1400--1403, 1981.

\bibitem{Gollub}
Y.~Sawada, A.~Dougherty, and J.~P. Gollub.
\newblock {\em Phys. Rev. Lett.}, 56(12):1260--1263, 1986.

\bibitem{Glicksman}
J.~S. Langer and H.~M\"uller-Krumbhaar.
\newblock {\em Phys. Rev. A}, 27(1):499--514, 1983.

\bibitem{Howison86}
S.~D. Howison.
\newblock {\em SIAM J. Appl. Math.}, 46(1):20--26, 1986.

\bibitem{KruSeg}
M.~D. Kruskal and H.~Segur.
\newblock {\em Stud. Appl. Math.}, 85(2):129--181, 1991.

\bibitem{Langer}
J.~S. Langer.
\newblock {\em Phys. Rev. Lett.}, 44(15):1023--1026, 1980.

\bibitem{Nakaya}
U.~Nakaya.
\newblock {\em Snow Crystals}.
\newblock Harvard University Press., Cambridge, 1954.

\bibitem{Richardson72}
S.~Richardson.
\newblock {\em Journal of Fluid Mechanics}, 56:609--618, 1972.

\bibitem{Varchenko/Etingof:1992}
A.N. Varchenko and P.I. Etingof.
\newblock {\em Why the boundary of a round drop becomes a curve of order four}.
\newblock American Mathematical Society, 1992.


\bibitem{Ed_Ferenc_Razvan}
F~Balogh, E~Saff, and R~Teodorescu.
\newblock in preparation.

\bibitem{Hohlov-Howison94}
Y.~E. Hohlov and S.~D. Howison.
\newblock {\em Quart. Appl. Math.}, 51(4):777--789, 1993.

 \bibitem{WT}
S-Y. Lee, R Teodorescu, and P Wiegmann, LA-UR 07-3622, unpublished.

\bibitem{Sw}
E.~Sharon, M.~G. Moore, W.~D. McCormick, and H.~L. Swinney.
\newblock {\em Phys. Rev. Lett.}, 91(20):205504, 2003.



\end{thebibliography}
\end{document}